\def\be{\begin{equation}}
\def\ee{\end{equation}}
\def\ba{\begin{eqnarray}}
\def\nn{\nonumber}
\def\ea{\end{eqnarray}}

\def\d{\delta}

\def\f{\phi}

\def\m{\mu}
\def\n{\nu}
\def\o{\omega}
\def\p{\pi}

\def\s{\sigma}
\def\t{\tau}

\def\O{\Omega}

\def\in{\infty}
\def\nt{\tilde{\nu}}
\def\mt{\tilde{\mu}}

\def\inti{\int_{-\infty}^{\tau_{i}(T)}}
\def\intij{\int_{-\infty}^{\tau_{j}(t_i(\tau_i))}}

\def\intiz{\int_{0}^{\infty }}

\newskip\humongous \humongous=0pt plus 1000pt minus 1000pt

\newif\ifdtup

\def\ha{{1\over 2}}

\def\(#1){(\ref{#1})}

\documentstyle[12pt]{article}

\makeatletter                    
\@addtoreset{equation}{section}  
\makeatother                     

\begin{document}

\title{Coherence and Fluctuations in the Interaction between Moving Atoms 
and a Quantum Field
\thanks{
Invited talk by B. L. Hu at the International Conference on Quantum Coherence,
Northeastern University, Boston, July 11-13, 1997.
To appear in the Proceedings edited by J. Swain and A. Widom
(World Scientific, Singapore, 1998).}}
\author{B. L. Hu\\
{\small Department of Physics, University of Maryland,
 College Park, Maryland 20742}\\
Alpan Raval\\
{\small Department of Physics, University of Wisconsin-Milwaukee,
Milwaukee, Wisconsin 53201 }}
\date{\small {\it (UMDPP 98-51, October 24, 1997)}}                  
\maketitle
\begin{abstract}
Mesoscopic physics deals with three fundamental issues: quantum coherence,
fluctuations and correlations. Here we analyze these issues for atom optics,
using a simplified model of an assembly of atoms
(or detectors, which are particles with some internal degree of freedom)
moving in arbitrary trajectories in a quantum field.  Employing 
the influence functional formalism, we study the self-consistent effect of the
field on the atoms, and their mutual interactions via coupling to the field.
We derive the coupled Langevin equations for the atom assemblage and
analyze the relation of dissipative dynamics of the atoms (detectors) with the
correlation and fluctuations of the quantum field. This provides a useful
theoretical framework for analysing the coherent properties
of atom-field systems. 
\end{abstract}

\newpage
\section{Mesoscopic Physics in Condensed Matter, Atom Optics, and Cosmology}

To practitioners in condensed matter physics, mesoscopia
refers to rather specific problems where, for example, the sample size is
comparable to the probing scale (nanometers), or the interaction time is
comparable to the time of measurement (femtoseconds), or that the electron
wavefunction is correlated over the sample thus changing its transport
properties fundamentally, or that the fluctuation pattern
is reproducible and sample specific. In atom/radiation optics, it is the
regime where coherent atom-field interaction,  correlations of field,
or the effect of boundaries become important. In cosmology, it is the
epoch when quantum fluctuations of fields mediate phase transitions, reheat
the universe or seed the galaxies. They are described by semiclassical gravity
and unified theories from the Planck to the GUT scales.
Here we work with a generalized definition
of mesoscopia proposed by one of us (see \cite{meso}, where a general
discussion
of the conceptual unity among  these disciplines can be found), i.e.,
the quantum / classical, micro / macro interface. It also entails
coherent / decoherent,  stochastic / deterministic dynamics, and
discrete / continuum correspondences.
As pointed out in \cite{meso}, mesoscopia deals with three fundamental
issues: quantum coherence, fluctuations and correlations. All mesoscopic
processes involve one or more of these aspects.

Many current research directions in early universe cosmology and
black hole physics also involve these aspects in a fundamental way.
The focus of this talk is however exclusively on atom / radiation optics,
which deals with the coherent interaction of atoms and radiation.
We will consider the interaction of an atom with a quantum field and
examine the coherence, correlation and fluctuations of such a system
in a fully non-equilibrium, relativistic field-theoretical treatment \cite{RHA}
\footnote{This treatment is more than
necessary for atom optics which deals with slowly moving atoms, and perhaps
more befitting for fast moving charged particles in strong fields
(plasma physics), but the consistency of backreaction makes such a demand,
and it is safer to take the finite temperature, nonrelativistic, far-field,
slow motion limits from the final result than as simplifying conditions
ab initio. See \cite{JohHu}.}. This situation is of basic interest
because quantum fields possess zero-point
fluctuations which manifest as random forces on an atom. The coherence
of the vacuum state also enters in an essential way in the description
of atom-field interactions. Here, we shall use a simplified model of an
atom, that of
a particle with an internal oscillator degree of freedom -- call it a
detector,
moving along an arbitrary trajectory. In fact, to make the correlation
aspects even more manifest, we consider an assembly of n such detectors
coupled
to a quantum field, and study their interaction with the field and their
mutual interactions via coupling to the field. This type of
problem has been treated before when the atoms (or detectors)
are stationary. In \cite{louis}, for example, there is a
discussion of Langevin equations for an arbitrary number of
homogenously broadened three-level atoms. Such a treatment, however, holds
for atoms fixed in space, and does 
not consider arbitrary states of motion of the atoms
themselves. 

On the other hand, nowhere is the role of  fluctuations of the vacuum 
more explicit than in the motion of a uniformly accelerated atom or
detector. In the
frame moving with such a detector, fluctuations of different modes of
the vacuum combine so as to appear as  thermal fluctuations with a
temperature
proportional to the acceleration (the Unruh effect \cite{Unr}).
Although this effect is considered too small to be directly measurable
via mesoscopic experiments at the present time (see \cite{rosu} for a
discussion of
numerical estimates on the acceleration for a measurable Unruh effect), it
does lead to shifts in the 
energy levels of two-level atoms which are qualitatively of the same form
as the Lamb 
shift, and can therefore be ascribed to vacuum fluctuations. 
Atomic energy shifts resulting from accelerated motion, and their
origin in  vacuum fluctuations and radiation reaction have been investigated
\cite{AudMul}. It has
also been recently shown \cite{Marzlin} that acceleration induces new
Raman-like transitions in multi-level atoms.

Experiments
will soon reach the stage where such delicate field-theoretical and
statistical mechanical attributes of an atom-field system, 
and their sensitivity to different states of motion, 
 will become important and measurable. 
Thus it is instructive to examine the effect of introducing two types of
theoretical ``probes'' into the coherence and fluctuation properties of a
quantum state. One type of probe involves introducing a large number of
atoms and allowing the quantized field to mediate interactions between
them, thus setting up correlations in such a many-body system. The other
type of probe is to ascribe arbitrary states of motion to the atoms
themselves, thus amplifying vacuum fluctuations. It is highly likely that
consideration of such problems will lead to new ways  of understanding
phenomena associated with vacuum fluctuations and quantum coherence. 

We treat the simplest version of such a problem in the next section,
details of 
which can be found in \cite{RHA}. 
We find that the dynamics of a many detector system can be described by a
set of coupled Langevin equations, and that the noise sources in these
Langevin equations are governed by vacuum fluctuations of the field at the
site of the detector and by quantum correlations of the vacuum between
different detector sites. Also, as we have shown in \cite{RHA}, these
equations lead  to the following conclusions:\\
\noindent a) Vacuum fluctuations are related linearly (albeit nonlocally
in general) to the dissipative or radiation reaction self-force on the
atom by a fluctuation dissipation relation which holds for atoms on
arbitrary trajectories (not just stationary trajectories, as is usually
assumed in the literature -- see \cite{AudMul} for example). \\
\noindent b) The form of the dissipative radiation reaction force is
independent of the trajectory in our model (in \cite{AudMul} the same
statement is proved independently for stationary trajectories using a
realistic QED model, thus lending support to its general validity). \\
\noindent c) Finally, correlations of the vacuum between different
detector
sites are related linearly to the radiation mediated between them for
trajectories without horizons (all realistic particle trajectories fall in
this class). This leads us to generalizations of the fluctuation-
dissipation relation to ``correlation-propagation'' relations, explicitly 
displayed in \cite{RHA}.

All of the above three statements, if shown to hold for realistic atom
optics, are experimentally testable in principle and should be essential
components of a
non-equilibrium description of mesoscopic atom optics systems. We will now
go on to outline the relevant details of our model. 

\section{Coherent Atom Optics: Moving Atoms in Scalar Electrodynamics Model}

We use a simplified model which describes the coupling of some
charged particles (we shall interchangeably call them atoms or
detectors) with some internal degrees of freedom
to a massless scalar field. Although realistic atoms are electrically
neutral, this type of interaction is similar to the interaction of the
atomic dipole moment with a electromagnetic field. To find out the effect
of the field on
the detectors moving along arbitrary trajectories, we derive the influence
functional, and from it the coupled Langevin equations of motion  for
the  system of $N$ detectors. The effect of correlation, dissipation and
fluctuations
can be extracted. We assume that the field and the system of
detectors are initially decoupled from each other, and that the field is
initially in the Minkowski vacuum state. We consider a 1+1 dimensional
model, even though the formalism can be simply extended
to higher dimensions, and to different choices of initial state for the
field, for example a finite temperature density matrix. The
path integral influence functional formalism we
use is somewhat different from the operator formalisms usually employed in
atom optics. It allows one to characterize fluctuation and dissipation
arising from the field variables which are integrated out, in a natural
way. It is also a self-consistent, non-perturbative treatment (the
coupling constant between atom and field need not be treated as a small
parameter when the Lagrangian is quadratic in all variables) and takes
into account the full backreaction of the field on the atoms.

Consider $N$ detectors or atoms $i=1,..N$ in $1+1$ dimensions with
internal 
coordinates $ Q_i(\tau_i)$ modeled as oscillator degrees of freedom,  and
moving on trajectories $(x_i(\tau_i), t_i(\tau_i))$
, $\tau_i$ being a parameter along the trajectory of detector $i$.
In the following analysis, we do not need to assume that $\t _i$ is the proper
time, although this is, in most cases, a convenient choice. However, we will
assume hereafter that the trajectories $(t_i(\tau_i),x_i(\tau_i))$
are smooth and that the parameters $\tau_i$ are chosen such that
 $t_i(\tau_i)$ is a strictly increasing function of $\tau_{i}$.

The detectors are coupled to a massless scalar field $\phi(x,t)$ via the
interaction action
\be
S_{int} = \sum _i e_i \int _{-\infty }^{t_i^{-1}(T)} d\tau_i   s_{i}(\tau_{i})
\frac{d Q_i}{d\tau_i} \phi (x_i(\tau_i), t_i(\tau_i)).
\ee
Here, $T$ is a global Minkowski time coordinate which defines a spacelike
hypersurface, $e_i$ denotes the coupling
constant of detector $i$ to the field, $s_{i}(\tau_{i})$ is a switching
function for detector $i$ (typically a step function), and $t_{i}^{-1}$
is the
inverse function of $t_{i}$. $t_i^{-1}(T)$ is therefore the value of $\t _i$ at
the point of intersection of the spacelike hypersurface defined by $T$ with the
trajectory of detector $i$. Note that the strictly increasing property of
$t_i(\t _i)$ implies that the inverse, if it exists, is unique.

The action of the system of detectors is
\be
S_{osc} = \ha \sum _i \int _{-\infty }^{t_i^{-1}(T)} d\tau_i
 [(\partial_{\tau_i}Q_{i})^2 - \O _{i}^2 Q_{i}^2].
\ee
The scalar field action is given by
\be
S_{field} = \frac{1}{2} \int_{-\infty }^{T} dt \int dx  [(\partial_t\f )^2 -
(\partial_x\f )^2]
\ee
and the complete action
\be
S = S_{field} + S_{osc} + S_{int} .
\ee

Expanding the field in normal modes,
\be
\phi(x,t) = \sqrt {\frac{2}{L}} \sum '_k [ q_k^+ (t) \cos  kx + q_k^- (t) \sin
kx]
\ee
where $\sum '_k$ denotes that the summation is restricted to the upper half
$k$ space, $k>0$.
Then the action for the scalar field is given by ($\sigma = +,-$)
\be
S_{field} = \frac{1}{2} \sum '_{k,\sigma} [ (\dot q_k^\sigma)^2 - \omega_k^2
q_k^2]
\ee
and the interaction action is
\ba
S_{int} &=& \sum _{i}e_i \sqrt {\frac{2}{L}} \int_{-\infty }^{t_i^{-1}(T)}
d\tau_i
\frac{dQ_i}{d\tau_i} \times     \nn \\
& &\sum '_{k} [ q_k^+ (t_i(\tau_i)) \cos  kx_i(\tau_i)
+ q_k^- (t_i(\tau_i)) \sin  kx_i(\tau_i) ] s_i(\tau_i)    \nn \\
       &=& \sum _{i}e_i \sqrt {\frac{2}{L}} \int_{-\infty }^{\infty }dt
        \int_{-\infty }^{t_i^{-1}(T)} d\tau_i \d (t-t_i(\tau_i))
        \frac{dQ_i}{d\tau_i} \times  \nn \\
        & &\sum '_{k} [ q_k^+ (t) \cos  kx_i(\tau_i) + q_k^- (t) \sin
kx_i(\tau_i) ]
        s_i(\tau_i).
\ea
We have $t_{i}(\tau_i) < T$, which follows from $\tau_i < t_i^{-1}(T)$ and
the property that $t_i(\tau_i)$ is a strictly increasing function. Hence
we may replace the upper limit of the $dt$ integration by $T$. This
replacement leads to the expression:
\be
S_{int} = - \sum '_{k,\sigma} \int_{-\infty }^{T} dt J_k^\sigma (t)
q_k^\sigma (t)
\ee
where
\be
J_k^{\sigma}(t) = -\sum_i e_i \sqrt{\frac{2}{L}} \int_{-\in }^{t_i^{-1}(T)}
d\tau_i
\delta(t-t_i(\tau_i)) \frac{dQ_i}{d\tau_i} u_{k}^{\sigma}(\tau_i) s_i(\tau_i)
\ee
and
\be
u_{k}^+(\tau_i) =  \cos  kx_i(\tau_i) ;~~~u_{k}^-(\tau_i) =  \sin
kx_i(\tau_i).
\ee

The action $S_{field} + S_{int}$ therefore describes a system of decoupled
harmonic oscillators each driven by separate source terms. The zero
temperature influence functional (corresponding to the initial state of the
field being the Minkowski vacuum state) for this system has the form
\cite{HPZ}:
\be
{\cal F}[J,J'] = \exp - {1\over \hbar} \sum '_{k,\sigma} \int_{-\in }^T ds
\int_{-\in }^s ds'[J_k^{\sigma}(s)-J_k^{'\sigma}(s)] [\zeta_k (s,s')
J_k^{\sigma}(s')
 -\zeta_k^*(s,s')J_k^{'\sigma} (s')] 
\ee
where
\be
\zeta_k \equiv \nu_k + i\mu_k = \frac{1}{2\omega_k}e^{-i\omega_k(s-s')}.
\ee
If the field is initially in a thermal state, the influence functional has
the same form as above, and the quantity $\zeta_k $ becomes
\be
\zeta_k = \frac{1}{2\omega_k}\left[\coth\left(\ha \beta \o
_k\hbar\right)\cos\omega_k(s-s')
-i\sin\omega_k(s-s')\right],
\ee
$\beta$ being the inverse temperature.
We shall restrict our attention to the zero temperature case.

Substituting for the $J_k^{\sigma}$'s in the influence functional, and carrying
out the $\delta$-function integrations, one obtains

\ba
{\cal F}[\{Q\};\{Q'\}] &=& \exp -{1\over \hbar}\left\{ \sum_{i,j=1}^{N}
\int_{-\in
}^
{t_i^{-1}(T)}d\tau_i s_i(\tau_i) \int_{-\in }^{t_j^{-1}(t_i(\tau_i))} d\tau'_j
s_j(\tau'_j) \left[\frac{dQ_i }{d\tau_i}-\frac{dQ'_i}{d\tau_i}\right]
\times \right.\nn \\
& &\left.\left[ Z_{ij}  (\tau_i, \tau '_j) \frac{dQ_j}{d\tau '_j}
- Z_{ij}^*(\tau_i, \tau '_j) \frac{dQ'_j}{d\tau '_j} \right]\right\}
\ea
where
\be
 Z_{ij}(\tau_i, \tau '_j) =\frac{2}{L} e_i e_j \sum '_{k,\sigma}
\zeta_k(t_i(\tau_i),t_j(\tau'_j)) u_{k}^\sigma (\tau_i) u_{k}^\sigma (\tau
'_j).
\ee
In the above, the continuum limit in the mode sum is recovered through
the replacement $\sum '_k \rightarrow \frac{L}{2\p } \int_0^{\in }dk$. We then
obtain, after substituting for $u_k^{\sigma }$ and $\zeta_k$,
\be
 Z_{ij}(\tau_i, \tau '_j) =\frac{e_i e_j}{2\p } \int_0^{\infty } \frac{dk}{k}
e^{-ik(t_i(\tau_i)-t_j(\tau'_j))} \cos k(x_i(\tau_i) - x_j(\tau'_j)).
\ee
In this form, $Z_{ij}$ is proportional to the two point function of the
free scalar field in the Minkowski vacuum, evaluated for the two points
lying on trajectories $i$ and $j$ of the detector system. It obeys the
symmetry relation
\be
 Z_{ij}(\tau_i, \tau '_j) =  Z_{ji}^{\ast }(\tau'_j, \tau _i)
\ee

Corresponding to ($2.12$), we may also split $Z_{ij}$ into its real and
imaginary parts. Thus we define
\be
Z_{ij}(\t _i,\t '_j) = \tilde{\n }_{ij}(\t _i,\t '_j) + i\tilde{\m }_{ij}(\t
_i,\t '_j)
\ee
where
\ba
\nt _{ij}(\t _i,\t '_j) &=& \frac{e_i e_j}{2\p } \int_0^{\in } \frac{dk}{k}
\cos k(t_i(\tau_i)-t_j(\tau'_j)) \cos k(x_i(\tau_i) - x_j(\tau'_j))   \nn \\
\mt _{ij}(\t _i,\t '_j) &=& -\frac{e_i e_j}{2\p } \int_0^{\in } \frac{dk}{k}
\sin k(t_i(\tau_i)-t_j(\tau'_j)) \cos k(x_i(\tau_i) - x_j(\tau'_j)).
\ea
$\nt $ and $\mt $ are proportional to the anticommutator and the
commutator of the field in the Minkowski vacuum, respectively.

The quantities $Z_{ij}$ are also conveniently expressed in terms of
advanced and retarded null coordinates $v_i(\t _i) = t_i(\t _i) + x_i(\t _i)$
and $u_i(\t _i) = t_i(\t _i) - x_i(\t _i)$, as
\be
Z_{ij}(\t _i,\t '_j) = Z_{ij}^{a}(\t _i,\t '_j) + Z_{ij}^{r}(\t _i,\t '_j)
\ee
where
\ba
Z_{ij}^{a}(\t _i,\t '_j) &=& \frac{e_ie_j}{4\p }\intiz \frac{dk}{k} e^{-ik(v_i
(\t _i)-v_j(\t '_j))}  \nn \\
Z_{ij}^{r}(\t _i,\t '_j) &=& \frac{e_ie_j}{4\p }\intiz \frac{dk}{k} e^{-ik(u_i
(\t _i)-u_j(\t '_j))}
\ea
and the superscripts $a$ and $r$ denote advanced and retarded
respectively.\footnote{
The terminology `advanced' and `retarded' refers to the null coordinates.
Equivalently, they can be called `left-moving' and `right-moving',
respectively,
when the sense of motion refers to the future direction in time. This
terminology
is used in wave theory and string theory.}
Similar decompositions for $\nt _{ij}$ and $\mt _{ij}$ thus follow.

The influence functional, together with the free action for the detector
system,
can be employed to obtain the propagator for the reduced density matrix of
the
system of detectors. This propogator will contain complete information
about the dynamics of the detectors. However, we shall take the alternative
route of deriving Langevin equations for the detector system in order
to describe its dynamics.

\section{Stochastic Dynamics of Atom-Field Interactions: Dissipation,
Fluctuations and Correlations}

So  far,  we have shown how to
coarse-grain the field degrees of freedom and incorporate their
effect on the detectors which manifest as long-range interactions
between the various detectors. We now derive the effective stochastic
equations of motion for the $N$-detector system. These equations should be
considered strictly useful only after the system of detectors has
effectively
decohered and a consistent classical description then becomes valid. Only
in this regime can  stochastic variables replace the true quantum
variables to good approximation. 
Physically, one is usually interested in looking at the system long after
its transient behavior is damped out (i.e. after the relaxation time), and
in this regime the effective stochastic description is quite
sound\footnote{The stochastic description should be sound even over
time
scales much shorter than the relaxation time scale, but longer than the
decoherence time scale. However, the Langevin equations derived here
cannot be used to study the process of decoherence itself.}. The
reason we prefer such a description is that it allows us to characterize
quantum noise from the field as a bonafide stochastic variable, thus
making possible a natural probabilistic interpretation for the quantum
fluctuation-dissipation relation.

Going back to the form ($2.11$) for the influence functional, we define
the centre of mass and relative variables
\ba
J_k^{+\s }(s) &=& (J_k^{\s }(s) + J_k^{'\s }(s))/2   \nn \\
J_k^{-\s }(s) &=& J_k^{\s }(s) - J_k^{'\s }(s).
\ea
Correspondingly, we also find it convenient to define
\ba
Q^+_{i}(\t _i) &=& (Q_i(\t _i) + Q' _i(\t _i))/2   \nn \\
Q^-_{i}(\t _i) &=& Q_i(\t _i) - Q' _i(\t _i).
\ea
Then Equation ($2.11$) yields
\ba
\mid {\cal F}[J,J']\mid  &=& \exp \{- {1\over \hbar} \sum '_{k,\sigma}
 \int_{-\in }^T ds \int_{-\in }^s ds'J_k^{-\sigma}(s) \nu_k (s,s')
 J_k^{-\sigma}(s')\}       \\
 &=& \int \Pi '_{k,\sigma } ({\cal D}\xi_k^{\s } P[\xi_k^{\s }])
 \exp -{i\over \hbar} \sum '_{k,\sigma} \int_{-\in }^T ds J_k^{-\s }(s)
 \xi_k^{\s }(s).
\ea
$\mid {\cal F} \mid$ is the absolute value of ${\cal F}$, containing the
kernel $\nu_k$. The phase of ${\cal F}$ contains the kernel $\mu_k$.
In the second equality, we have used a functional gaussian integral
identity, $P[\xi_k^{\s }]$ being the positive definite measure
\be
P[\xi_k^{\s }] = N\exp \{- {1\over 2\hbar} \int_{-\in }^T ds \int_{-\in }^T ds'
\xi_k^{\sigma}(s) \nu_k^{-1} (s,s') \xi_k^{\sigma}(s')\}       \\
\ee
normalized to unity. It can therefore be interpreted as a probability
distribution over the function space $\xi_k^{\s }$.

The influence functional can thus be expressed as
\ba
{\cal F}[\{Q\},\{Q'\}] &=& < \exp \left\{- {i\over \hbar} \sum '_{k,\s }
\int_{-\in
}^T ds
J_k^{-\s }(s) \left[\xi_k^{\s }(s) + 2\int_{-\in }^{s} ds'\mu_k(s,s')
J_{k}^{+\s }(s')\right]\right\} >             \nn \\
&\equiv & < \exp {i\over \hbar} S_{inf} >
\ea
where $<\mbox{ }>$ denotes expectation value with respect to the joint
distribution $\Pi'_{k,\s }P[\xi_k^{\s }]$. $S_{inf}$ will be called the
stochastic influence action. We find
\ba
< \xi_k^{\s }(s) > &=& 0,     \nn \\
< \{\xi_k^{\s }(s), \xi_{k'}^{\s '}(s')\} > &=& \hbar \d _{kk'}\d _{\s \s '}
\nu_k(s,s')
\ea
where $\{\mbox{ }, \}$ denotes the anticommutator.

Substituting for $J_k^{-\s }$ and $J_k^{+\s }$ in terms of the detector
degrees of freedom $\{ Q_i \}$, the stochastic influence action $S_{inf}$ is
obtained as
\be
S_{inf} = -\sum_{i=1}^{N} \inti d\t _i \frac{dQ^- _i}{d\t _i} s_i(\t _i)
\left[\eta_i(\t _i) + 2\sum_{j=1}^{N} \intij d\t '_j \frac{dQ^+_j}{d\t
'_j} s_j(\t
'_j)
\mt _{ij}(\t _i,\t '_j) \right]
\ee
with
\be
\eta_{i}(\t _i) = e_i \sum '_{k,\s } \sqrt {2\over L} u_k^{\s }(\t _i)
\xi_k^{\s }
(t_i(\t _i)).
\ee
 From Equation ($3.8$) we see that the quantities $\mt _{ij}$, $i\neq j$
mediate
long-range interactions between the various detectors and
 the quantities $\mt _{ii}$
describe self-interaction of each detector due to its interaction
with the field. This self-interaction
typically manifests itself as a dissipative (or radiation reaction)
force in the dynamics of the detectors. We will, therefore,
refer to $\mt _{ij}$, $i\neq j$ as a ``propagation kernel'', and $\mt _{ii}$
as a ``dissipation kernel''.

We now turn to the interpretation of the quantities $\eta_i$. They appear as
source terms in the effective action of the detector system. Also, being
linear combinations of the quantities $\xi_k^{\s }$, they are stochastic
in nature. Indeed, from Equations ($3.7$) and ($3.9$) we can obtain
\ba
< \eta_i(\t _i) > &=& 0,     \nn \\
< \{\eta_i(\t _i), \eta_j(\t '_j)\}> &=& e_i e_j \sum '_{k,\s } \sum '_{k',\s
'}
u_k^{\s }(\t _i)u_{k'}^{\s '}(\t '_j) (\frac{2}{L})
< \xi_k^{\s } (t_i(\t _i)) \xi_{k'}^{\s '} (t_j(\t '_j))>   \nn \\
&=& \hbar \nt _{ij}(\t _i,\t '_j).
\ea
Thus $\nt _{ij}$ appears as a correlator of the stochastic forces $\eta_i$ and
$\eta_j$. Along a fixed trajectory, this correlation manifests as noise
in the detector dynamics. Hence we call $\nt _{ii}$ a ``noise kernel'' and
$\nt _{ij}$, $i\neq j$, a ``correlation kernel''.\footnote{The distinction
between noise and correlation is unnecessary from the point of view of the
field. `Noise', as used here, also represents free field correlations for
points on a single trajectory. However, from the point of view of each
detector, these two quantities play a different role. Hence the choice of
terminology.}

The full stochastic effective action for the $N$-detector system
is given by
\be
S_{eff} = S_{osc} + S_{inf}.
\ee
We may now express this in terms of the variables $Q^+_i$ and $Q^-_i$ defined
earlier. Thus we obtain
\ba
S_{eff} &=& \sum_{i=1}^{N} \inti d\t _i [ \dot{Q^- _i} \dot{Q^+_i} -
\O _{i}^{2} Q^- _iQ^+_i
- \dot{Q^- _i}s_i(\t _i)\eta_i(\t _i) \nn \\
& &- 2 \dot{Q^- _i}
s_i(\t _i) \sum_{j=1}^{N} \intij d\t '_j \dot{Q^+_{j'}} s_j(\t '_j)
\mt _{ij}(\t _i,\t '_j) ]
\ea
where $\dot{f_i} \equiv \frac{df_i}{d\t _i}$, $\dot{f_{j'}} \equiv \frac
{df_{j}}{d\t '_j}$.

Extremizing the effective action with respect to $Q^- _i$ and setting
$Q_i = Q'_i$ at the end \cite{HPZ}, we obtain a set of coupled
equations of motion, the Langevin equations, for the system of detectors:
\be
\frac{d^2 Q_i}{d\t _i^{2}} - 2\sum_{j=1}^{N} \intij d\t '_j s_j(\t '_j)
\frac{d}{d\t _i} (s_i(\t _i)\mt _{ij}(\t _i,\t '_j) ) \frac{dQ_j}{d\t '_j}
+ \O _i^2 Q_i = \frac{d}{d\t _i}(s_i(\t _i) \eta_i(\t _i)).
\ee
Due to the back-reaction of each detector on the field, and consequently on
other detectors,
the effective dynamics of the detector system is highly non-trivial
and, as such, can be solved in closed form only for simple trajectories
or under simplifying assumptions such as ignoring the back-reaction of
certain detectors on the field. For instance, if we choose to ignore the
back-reaction of detector $i$ on the field, this can be effected by setting
$\mt _{ji}=0$, for all $j$, including $j=i$,while at the same time keeping
$\mt _{ij} \neq 0$ for $j\neq i$. The particular case $\mt _{ii}=0$ amounts to
ignoring the radiation reaction of detector $i$. This is necessary because the
radiation reaction effect arises due to a modification of the field in the
vicinity of the detector as a consequence of the back-reaction of the detector
on the field\footnote{Of course, it is in general inconsistent to ignore 
the back-reaction
of a detector, as it leads to a direct violation of the symmetry ($2.17$).
As is well-known, it also leads to unphysical predictions. For example,
in the treatment of an atom on an inertial trajectory, coupled to a
quantum field, balance of vacuum fluctuations and radiation
reaction is necessary to ensure the stability of the ground state.
As explained above, ignoring back-reaction implies ignoring the
radiation reaction force. Such a treatment would render the ground
state unstable.
However, in certain cases, the quantities $\mt _{ji}$ may not contribute
to the dynamics of detector $j$. This occurs, for example, when
the
trajectory of one detector is always outside the causal future of the
other one. Hence there is no retarded effect of one of the detectors on
the other.}.

Our formal treatment of the detector-field system is exact in that it
includes the full back-reaction of the detectors on the field, which
is manifested in the coupled Langevin equations of the various detectors.
These coupled equations of motion give rise to a sort of ``dynamical
correlation''
between the various detectors. Non-dynamical correlations also occur because
of the intrinsic correlations in the state of the field (Minkowski
vacuum).
These correlations are purely quantum-mechanical in origin, and they are
reflected in the correlators of the stochastic forces, $\nt _{ij}$.
Correlations between stochastic forces on different detectors
induce correlations between the coordinates $Q_{i}$ of different detectors.

As commented earlier, and shown explicitly in \cite{RHA}, our exact
treatment makes it possible to
demonstrate the existence of generalized fluctuation-dissipation relations
relating the fluctuations of the stochastic forces on the detectors to the
dissipative forces. We also discovered a related set of correlation-propagation
relations between the correlations of stochastic forces on different
detectors and the retarded and advanced parts of the radiation mediated
by them. These relations are of categorical nature and hence of
fundamental significance in the desciption of moving atoms interacting
with a field.

\section{Applications}

As applications of these Langevin equations, we have considered in
\cite{RHA} four examples
of increasing complexity:
a) a single detector in the Minkowski vacuum moving on an inertial trajectory,
b) a single detector on a uniformly accelerated trajectory,
c) two detectors on inertial trajectories, and d) one detector on
 a uniformly accelerated trajectory
and another one on an arbitrary trajectory, functioning as a probe.
In all cases, we can solve exactly for the
detector coordinates, at least in the late time limit (this limit is actually
realized at any finite time when the two detectors have
been switched on forever, and corresponds to the neglect of transients in
the solutions for the detector coordinates).

Case a) describes the well-known physical effects in quantum field theory
of the dressing of a particle by the field. Case b) describes the Unruh effect
\cite{Unr} where thermal radiation from the excitation of quantum noise
is experienced by a uniformly accelerated particle.
In c) we introduce the notions of ``self'' and ``mutual'' impedance which govern
the response of either detector. The
effect of the back-reaction of each detector on the field and consequently
on the other detector is to introduce the so-called mutual impedance in
the detector response as well as to modify the self-impedance of each detector
from its value in the absence of the other one.
In d) we switch on the probe after it intersects the future horizon of the
uniformly accelerated detector, so that it cannot causally influence the
uniformly accelerated one. Because of this, the dissipative features of
this problem are relatively trivial. The response of the probe has contribution
mainly from field correlations across the horizon. On the other hand,
the noise due to field fluctuations and the field correlations
between the two trajectories play a dominant role. This correlation can
be expressed in terms of noise via a correlation-propagation relations
which are appropriate extensions of a generalized
fluctuation-dissipation relation directly relating field fluctuations
to dissipative properties of the detectors.


Here we have focussed on the correlation and fluctuation aspects of the
atom-field system. In more practical problems in atom optics, one can
use a suitable generalization of the Langevin equations derived here for
the description of dissipative
atom motion (with radiative-reaction).  The noise correlators  describe
the stochastic source from the vacuum fluctuations of the field as they
appear to moving atoms, and the fluctuation-dissipation
and correlation-propagation relations relate the dissipative effect of
the atom to the correlations and fluctuations of the field. Similar
methods are now applied to two-level atom systems \cite{AnaHu} and
detector motion with non-prescribed trajectories
(determined by  backreaction of the field) \cite{JohHu}. It is hoped that
studies such as these will bring new insights into the coherence and
fluctuation properties of the
quantum vacuum state, and further confirm the relevance of these
properties to mesoscopic
physics via system-field interactions.\\

{\bf Acknowledgement}  This research is supported in part by the National
Science Foundation under grants No. PHY94-21849 and PHY95-07740.

\end{document}